\documentclass[12pt]{article}
\def\be{\begin{equation}} \def\ee{\end{equation}}
\def\bea{\begin{eqnarray}} \def\eea{\end{eqnarray}} \def\ba{\begin{array}}
\def\ea{\end{array}} \def\ben{\begin{enumerate}} \def\een{\end{enumerate}}

\newcommand{\eqn}[1]{(\ref{#1})}

\newcommand{\plb}[3]{Phys. Lett. {\bf B#1} ({#2}) {#3}}

\newcommand{\hepth}[1]{{\tt hep-th/{#1}}}

\def\l{\lambda}
\def\m{\mu}

\def\n{\nu}

\def\br{\nonumber\\}

\begin{document}
{}~
\hfill\vbox{\hbox{hep-th/yymm.nnnn} \hbox{\today}}\break

\vskip 3.5cm
\centerline{\large \bf
 Galilean type IIA backgrounds and a map}
\vskip .5cm

\vspace*{.5cm}

\centerline{  
Harvendra Singh
}

\vspace*{.25cm}
\centerline{ \it  Theory Division, Saha Institute of Nuclear Physics} 
\centerline{ \it  1/AF Bidhannagar, Kolkata 700064, India}
\vspace*{.25cm}

\vspace*{.5cm}

\vskip.5cm
\centerline{E-mail: h.singh [AT] saha.ac.in }

\vskip1cm

\centerline{\bf Abstract} \bigskip

We study  non-relativistic 
 $AdS_4\times CP^3$ solutions 
with dynamical exponent 3 in  type 
IIA string theory, both with and without Romans mass. The  
 compactifications to four dimensions
are found to describe Proca fields in  anti-de 
Sitter spacetime. This leads us to conclude that the massive and massless 
IIA theories should be identified
 in four dimensions and the Romans' mass should be  
identified with the `flux' along $CP^3$. From 
supergravity point of view, it is 
suggestive  of a four-dimensional symmetry 
that rotates Romans mass into the flux along $CP^3$.  
We also identify M-theory Galilean (ABJM) background which gives rise to 
the nonrelativistic type IIA solution.
\vfill 
\eject

\baselineskip=16.2pt


\section{Introduction}
Recent applications of AdS/CFT holography [1,2,3] to strongly
coupled non-relativistic systems, 
showing scaling behaviour near quantum critical points, have been a
subject 
of wide attention \cite{son}-\cite{nakayama}. 
For studying the behaviour near quantum critical points one
considers
the nonrelativistic case of AdS/CFT holography which  exhibits  
a reduced conformal
symmetry [4,5] or  `Schroedinger group' \cite{mehen,mehen2,mehen3}. Particularly
holographic 
study of strongly coupled fermionic
systems at finite density has been termed as `AdS/Atoms' 
[4,5]. The  Galilean symmetries in physical systems have been
studied even earlier  \cite{horvathy,horvathy2}.  For the study
of finite
temperature properties like phase transitions, transport and viscosity
etc, one 
 includes black holes in the AdS backgrounds [6,7].
In  parallel studies of superconductivity under the probe 
approximations, the bulk
 AdS geometry generically involves
spontaneously broken (Higgs) phases where the Abelian field
coupled to a complex scalar field becomes massive \cite{herzogrev,denefrev}.

By now there have been 
quite a few explicit examples where  non-relativistic anti-de-Sitter 
(NRadS) geometries
could be embedded in type II string theory and
M-theory, see  \cite{marte,varela,donos,arnab,donos1,donos2a,
gaunt,gubser,hs,colgain}.
Also recently there have been some
attempts to obtain non-relativistic solutions in 
the  massive type IIA supergravity of Romans \cite{roma}.
The Romans theory is the only known example of a
10-dimensional
maximal supergravity where $B_{\m\n}$ field is explicitly massive to
begin with. There is also a cosmological constant in the theory.
Thus massive type IIA sugra provides an unique case to study  
interesting
NRadS solutions and the  holographic dual Galilean field theory.
In a recent work \cite{hs}  we specifically 
obtained a non-relativistic $AdS_4\times M^6$ solution, 
where $M^6$ is a Einstein space, e.g. $CP^3$, $S^3\times S^3$ or 
$S^6$. It is  
intriguing to ask what is its relationship with the $AdS_4\times CP^3$ 
ABJM backgrounds of ordinary type IIA \cite{abjm}, and its 
Galilean generalisation which we find in this paper. To recall, there 
also exists this fact for long, that the type IIA can be lifted to 
M-theory while there is no 11-dimensional analogue for the Romans 
massive type IIA theory. However, the two theories when compactified to 
lower 
dimensions could be mapped into each other via T-dualities \cite{hs1,hs2002,hs2}, 
 by switching on 
appropriate fluxes. We would like to see if there does exist such a map 
for the Galilean solutions in the two theories.

In this work we want to discuss some Galilean examples involving $CP^3$ 
compactifications and 
explore the relationship between ordinary type IIA and its
only known massive cousin. In  section-2 we construct Galilean 
solution of type 
IIA with dynamical exponent 3. We identify this as a 
non-relativistic generalisation of the ABJM solution. We also discuss its 
M-theory 
origin and also find  4-dimensional compactified effective action. In  
section-3 we review  previously known Galilean solution in Romans' theory 
which has  almost identical features. The section-4 contains the map 
involving Galilean solutions of 
ordinary type IIA and the Romans theory. We summarize in section-5.

\section{Galilean type-IIA solution}

We look for a non-relativistic generalisation of the ABJM solution 
\cite{abjm} of type 
IIA string theory.
To solve the equations of motion of type IIA theory  
we take the following non-relativistic ansatz for $AdS_4\times CP^3$ 
metric (Einstein frame) and fields  (particularly with dynamical 
exponent 3) 
\bea\label{sol2}
&&ds^2_{IIA}=\sqrt{R^3k}\left({1\over4}( -{2\beta^2(dx^{+})^2\over z^{6}} 
+{-2dx^{+}dx^{-}+dy^2+dz^2\over
z^2})  + ds^2_{CP^3} \right) ,\br
&&e^{\phi}={R\over k}, ~~~~
F_{+-yz}= {3\over 8}{ R^2 k\over z^4} \ , \br
&& B_{+ y}= \beta{ R^2p\over z^4}, ~~~C= \beta {qk\over z^3}dx^{+} +k 
\omega 
\eea 
where $ J=d \omega$ defines the K\"ahler 2-form 
over $CP^3$.
These Galilean solutions exist provided   
$$q=2p={1\over \sqrt{2}} \ .$$
The string coupling is fixed by the ratio ${R\over k}\equiv g_o$. 
The parameters $R$ and $k$ have 
interpretation as in 
the ABJM 
work \cite{abjm}. That is, $R$ is the measure of the 
radius of 
$CP^3$ in the string frame $(\alpha'=1$) and $k$ is the quantum of 
2-form flux along $CP^3$. 
While in M-theory picture $k$ is the order of the orbifold $R^8/Z_k$. 
Since $k$ 
  fixes the string coupling, therefore 
$k$ must be taken sufficiently large to remain in the type IIA framework. 
The constant 
$\beta$ in the above is arbitrary  and can 
be easily scaled away. However we have kept it here because a relativistic 
solution 
is readily obtained by simply setting $\beta=0$ in \eqn{sol2}. 

The D-brane interpretation remains the same as of the ABJM except that we 
have 
got extra non-relativistic matter  fields $B_{+y}$ and $C_{+}$ 
contributing as  the `dust'. Due to this the $T_{++}$ 
component of energy-momentum tensor is nontrivial which otherwise 
would vanish in the relativistic case. 
Rest of the components of the 
energy momentum tensor stay as in the relativistic ABJM case. We also note 
that the transverse 
$CP^3$ metric is  
undeformed. 

Thus to summarise, there exists a non-relativistic $AdS_4\times CP^3$ 
background in type IIA string theory with the dynamical 
exponent being $3$. These solutions have Schr\"odinger defformation 
$(dx^{+})^2$ in the metric and have $B$ field. So these are more 
like the Schr\"odinger solutions (with dynamical exponent 2) of 
\cite{marte} and should be obtainable via TsT duality transformation. 
We shall comment about the supersymmetries in the next subsection.

\subsection{Galilean ABJM theory}
An M-theory lift of the type IIA Galilean solutions \eqn{sol2} can  be 
done and the corresponding 11-dimensional background is 
\bea\label{sol2m}
&&ds^2_{11}={{\hat R^2}\over4}( -2e^{+}e^{-}+e^ye^y+
e^z e^z)  + \hat R^2 ds^2_{CP^3}  +{\hat R^2\over 
k^2}e^\psi e^\psi,\br
&&
C_{3}={\hat R^3\over 8}e^{+}\wedge e^{-} \wedge e^y +  {\hat R^3\over 
2\sqrt{2}k}e^{+}\wedge e^{y} \wedge e^\psi
\eea
where the vielbeins are 
\be\label{viel1}
e^{+}={\beta dx^{+}\over z^3},~
e^{-}={z\over\beta} dx^{-}+{\beta dx^{+}\over z^3},~
e^{y}={dy\over z},~
e^{z}={dz\over z},~
e^\psi=d\psi+ { \beta k\over \sqrt{2}z^3}dx^{+}+ k\omega \ . \ee
In the above 
$R^2=\hat R^3/k$ and $\psi\sim\psi+2\pi$ is the 
11-th direction fibered over the base 
$ CP^3$. The $CP^3$ radius,
$\hat R$, is however  measured in 11-dimensional Planck length units.
The solution \eqn{sol2m}  describes a Galilean generalisation 
of the M-theory background \cite{abjm} corresponding to a stack of $N$ 
M2-branes 
placed on ${R^8\over Z_k}$ orbifold singularity. We especially mention that 
when  compared to the relativistic case ($\beta=0$) the coordinate
$\psi$ here appears to be twisted along $AdS_4$ as 
well as being fibered over the base $CP^3$.
(Note that one should not try to set $\beta=0$ directly 
into the vielbeins \eqn{viel1} instead it should be done at the 
level of the solution \eqn{sol2m}.) The solutions 
\eqn{sol2m} exist for  
arbitrary $k$ value. Fortunately these 11-dimensional solutions 
were 
already  constructed in \cite{varela} but  the relationship 
to ABJM work  was not
explored by the authors there. Our study makes this aspect  vividly 
clear. According to  \cite{varela} 
these 11-dimensional 
solutions do preserve two of the Poincar\'e 
supersymmetries
while all conformal supersymmetries are broken, see \cite{donos}  
\footnote{I am grateful to Oscar Varela for pointing out  error in 
the supersymmetry analysis and introducing the reference \cite{varela}.}

\subsection{Skew-whiffed case}
However, it has been shown in \cite{varela,donos} that  
the flipping of the sign of the 4-form flux
in  11-dimensional solution \eqn{sol2m} breaks supersymmetries completely.
Thus, a skew-whiffed 11-dimensional background can be written as 
\bea\label{sol2m1}
&&ds^2_{11}={{\hat R^2}\over4}( -2e^{+}e^{-}+e^ye^y+
e^z e^z)  + \hat R^2 ds^2_{CP^3}  +{\hat R^2\over 
k^2}e^\psi e^\psi,\br
&&
C_{3}=-{\hat R^3\over 8}e^{+}\wedge e^{-} \wedge e^y -  {\hat R^3\over 
2\sqrt{2}k}e^{+}\wedge e^{y} \wedge e^\psi
\eea
where the vielbeins are as in \eqn{viel1}. Corresponding to this, 
the type IIA  solution 
\eqn{sol2} will have the signs of $F_{+-yz}$  and $B_{+y}$ both negative
\bea\label{sol2e}
&&ds^2_{IIA}=\sqrt{R^3k}\left({1\over4}( -{2\beta^2(dx^{+})^2\over z^{6}} 
+{-2dx^{+}dx^{-}+dy^2+dz^2\over
z^2})  + ds^2_{CP^3} \right) ,\br
&&e^{\phi}={R\over k}, ~~~~
F_{+-yz}= -{3\over 8}{ R^2 k\over z^4} \ , \br
&& B_{+ y}= -\beta{ R^2\over 2\sqrt{2}z^4}, ~~~C= \beta {k\over\sqrt{2}
 z^3}dx^{+} +k \omega .
\eea 
We are specifically interested in  nonsupersymmetric Galilean
solution because the 
solution \eqn{sol2s} which we are going to compare it with 
is also nonsupersymmetric. 

\subsection{Compactification to $D=4$}
The $D=4$ truncation of the skew-whiffed 11-dimensional background 
 \eqn{sol2m1} can be performed consistently, see for details \cite{varela}. 
An effective action describing the compactification of the 11-dimensional
(nonsupersymmetric) solution \eqn{sol2m1} to  four  
dimensions is
\bea
\label{eq3a1}
S_4\sim\int d^{4}x \sqrt{-g}\bigg[ R
-{1\over 2.2!} (F_{\m\n})^2
-{1\over 2!} {12\over l_o^2}(A_{\m})^2 +{6\over l_o^2} \bigg]
\eea
alongwith the null conditions, which need to be imposed from outside,
 $F\wedge * F =A\wedge * A=0$. The action \eqn{eq3a1}
 describes a  vector field in anti-de Sitter spacetime with 
mass square as $12/l_o^2$. Note that $l_o^{2}= (R^3 k)^{1/2}/4$.  
The Proca action \eqn{eq3a1} admits a  Galilean solution 
\bea\label{sol21a}
&&ds^2=l_o^2\left(-{2 \over z^{6}}
(dx^{+})^2+{-2dx^{+}dx^{-}+dy^2+dz^2\over z^2} \right)  ,\br
&&
A_{+}=  {2\sqrt{2}\over \sqrt{3} }{l_o\over z^3}   
\eea
with dynamical exponent as 3, whose
 11-dimensional uplifted solution is \eqn{sol2m1}. 

All this  is quite similar to a  Galilean solution of massive type IIA
obtained in \cite{hs}, but 
with subtle differences. Namely, the ordinary type  IIA solution 
\eqn{sol2} and the skew-whiffed case are supported by 
2-form flux along $CP^3$, while the massive type IIA solution 
\eqn{sol2s} is devoid of the 
 2-form flux. So in the next section we first review the 
Galilean solution 
of massive type IIA supergravity for our comparative study.   

\section{A Galilean background in massive type IIA}

The Romans type IIA supergravity theory  in ten 
dimensions includes 
massive tensor field and  cosmological constant while being  maximally 
supersymmetric. It was  recently 
found in \cite{hs}
 that the theory admits  non-relativistic anti-de Sitter vacua, 
$NRadS_4\times CP^3 $, a Galilean solution  with dynamical exponent 
$3$, 
\bea\label{sol2s}
&&ds^2=L^2\left(-{2\beta^2\over z^{6}}
(dx^{+})^2+{-2dx^{+}dx^{-}+dy^2+dz^2\over
z^2}  +{5\over 2} ds_{CP^3}^2 \right) ,\br
&&e^{2\phi}={f^2\over m_0^2}, ~~F_{+-yz}= \sqrt{5} f {L^4\over z^4} \br
&&
B_{+ y}= \beta g_m^{1\over 2}{\sqrt{2} L^2\over z^4} , ~~~C_{+}= 
\beta g_m^{-{3\over 4}}{2\sqrt{5}L\over 3 
z^3} \ ,
\eea
where $L^{2}=2/(m_0^2~g_m^{5\over 2})$. In the above $m_0$ is the Romans 
mass 
 while $f$ is the measure of 4-form flux. The flux $f$  fixes the 
string 
coupling of the background. To distinguish it from $g_o$ we 
shall denote the string
coupling here as
$g_m\equiv f/m_0$. Note the  metric in \eqn{sol2s} is written in the 
Einstein frame. This Galilean
solution preserves no supersymmetry even in the  
relativistic case ($\beta=0$).

On compactification over $CP^3$,  a 4-dimensional 
effective action can be written as \cite{hs} 
\bea\label{eq3}
S_{4}&\sim & \int d^{4}x \sqrt{-g}\bigg[ R
-{1\over 2.3!}{1\over g_m}(H_{\m\n\l})^2
-{1\over 2.2!} g_m^{3\over2} (G_{\m\n})^2 +{6\over L^2}
\bigg] \br
&&-{\sqrt{5}m_0\over 2!2!}g_m^{3/2} \int
d^4x\epsilon^{\m\n\l\rho}(
B_{\m\n} G_{\l\rho}- {m_0\over 2}
B_{\m\n} B_{\l\rho}) \ ,
\eea
where $G_2\equiv dC_1+m_0 B$.
A simple exercise determines that if we integrate out   $G$ by
using its field equation, we simply obtain an action for the tensor field
\bea\label{eq4}
S_{4}&\sim &\int d^{4}x \sqrt{-g}\bigg[ R
-{1\over 2.3!}{1\over g_m}(H_{\m\n\l})^2
-{5m_0^2\over 2.2!}g_m^{3\over2} ( B_{\m\n})^2 +{6\over L^2}
\bigg] \br &&
+ {\sqrt{5}m_0^2\over 2 (2!)^2}g_m^{3\over 2} \int
d^4x \epsilon^{\m\n\l\rho}
B_{\m\n} B_{\l\rho} \ .
\eea
At this stage, we can  introduce a vector field
through a generalised Hodge-duality relation \cite{hs} in 4-dimensions as
\be\label{oi0}
\star H_3=  d\chi + \bar A_1,
\ee
where $\chi$ is the axion.  We have
introduced gauge field via  Hodge-dual relation in \eqn{oi0}, but 
in doing so the gauge field $\bar A$  actually 
 gauges the axionic shift symmetry.
The axion field serves as a
 Goldstone mode and
corresponding local shifts (Stueckelberg) are
\be
 \delta \chi=-\l,~~\delta \bar A_1=d\l .
\ee
This shift symmetry can eventually be used to set $\chi=0$.
Correspondingly the Proca action involving  $\bar A_\mu$ can be written 
as
\bea\label{eeq3a}
S_4\sim\int d^{4}x \sqrt{-g}\bigg[ R
-{1\over 2.2!} (\bar F_{\m\n})^2
-{1\over 2!}{12\over L^2}  (\bar A_{\m})^2 +{6\over L^2} \bigg]\ .
\eea
where $2L^{-2}=m_0^2 g_m{}^{5\over 2}$ as  given earlier. 
 
\section{Identification of Romans mass with type IIA  flux}

The two separate actions \eqn{eq3a1} and \eqn{eeq3a} describe the 
dynamics 
of  Proca fields coupled to anti-de Sitter gravity.  The former comes 
from 
ordinary type IIA  having fluxes along $CP^3$ while the latter comes 
from the $CP^3$ compactification of  Romans theory. Since we do not 
have
 any precise relationship between two ten-dimensional theories when 
$CP^3$ flux 
compactifications are involved, the similarity of the Proca actions 
\eqn{eq3a1} and 
\eqn{eeq3a} could be taken as a hint. So it will be 
appropriate to identify these two actions.
 
An identification between 4D actions 
\eqn{eq3a1} and 
\eqn{eeq3a} can be achieved by 
comparing the  
$AdS_4$ radii of  10-dimensional solutions \eqn{sol2e} and 
\eqn{sol2s}. That is we would, in fact, identify ${l_o^2}\equiv 
L^2$ hence
\be\label{map1}  
m_0^2 g_m{}^{5\over 2}\equiv{8\over \sqrt{R^3 k}} . \ee 
Thus we already see that the  mass parameter $m_0$ gets related 
to the 2-form flux $k$ of 
the type IIA  ABJM 
background in some manner. 
Note, however, that  in the Galilean solutions  the $AdS_4$ radius of 
curvature  is tied to  $CP^3$ 
radius in a definite way.  
The two $CP^3$ radii should then be related as
\be
({R_{m,CP^3}\over R_{o,CP^3}})^2={5\over 8}.
\ee
(The suffixes $m$ and $o$ are used in order to 
distinguish massive 
and ordinary type IIA cases.)  
However to be  precise we also need to relate the  
 Newton's constants in 4D actions \eqn{eq3a1} and 
\eqn{eeq3a}.\footnote{The $4D$ 
Newton's constant is
$G_4^N={G_{10}^N\over V_6}\propto {g_s^2\alpha'^4\over (R_{CP^3})^6}$, 
where $V_6={\pi^3\over 6} R_{CP^3}^6$ represents volume of a 
$CP^3$ manifold.} 
So if we identify  the 4-dimensional 
Newton's constants we  
find that 
10-dimensional string couplings must be related as
\be
g_m=c g_o
\ee 
with  $c=(5/8)^{3\over 2}$ being a numerical
constant. So we get
\be\label{map2}
  {f\over m_0} = c{R\over k} .
\ee
Following from Eqs.\eqn{map1} and \eqn{map2} a precise map can now be 
summarised as  
\be
f\leftrightarrow {2\sqrt{2}\over c^{5/4}}{1\over R}, ~~~~ 
m_0R\leftrightarrow {2\sqrt{2} \over c^{5/4}} {k\over R}.
\ee
This indicates that
 the  `mass' $m_0$  of Romans' background 
indeed gets mapped  to  the ABJM `flux' $k$ in
 ordinary type IIA solution.

In summary, it is interesting to have explicitly obtained  this 
relationship from
$CP^3$ compactifications especially involving Galilean 
non-supersymmetric solutions. But this may not be 
the first instance where such an equivalence has arisen. A somewhat similar 
type of situation has been reported in  
 recent works \cite{gaitt}
on ABJM theory with Romans mass. 
In these works there is an overall Chern-Simons level 
$$k=k_1+k_2\ne 0$$  
appearing in the ABJM theory. 
Actually it happens when the Chern-Simons levels in the product group
 $U(N)_{k_1}\times U(N)_{k_2}$ do not exactly cancel. The 
unbalanced flux $k$ deforms the original  ABJM 
theory. This setting  
also breaks all the supersymmetries \cite{gaitt,gaitt2}. Earlier too,  Romans 
mass  has been mapped through
 T-dualities into the 
fluxes on ordinary type IIA side, especially in  
toroidal \cite{berg, hs1} and $K3$ compactifications \cite{hs1,hs2002,hs2}, as 
well as 
 Calabi-Yau compactifications \cite{Louis:2002ny,louis2}. 

\noindent{\bf Symmetries:}\\
Further strong evidence in favour of our proposal is the nature 
of symmetries of two type IIA solutions we have discussed. 
The Galilean type IIA solutions \eqn{sol2} and 
\eqn{sol2e} 
have non-relativistic $AdS_4$ symmetries as well as they inherit global 
$SU(4)$ 
symmetry comming from round $CP^3$'s. We see that precisely
the same amount of symmetries are exhibited by the massive type IIA 
Galilean solutions
\eqn{sol2s} as well. 

\section{Conclusion}
We obtained a Galilean type IIA background as a direct deformation of the 
ABJM solution \cite{abjm} and in this way we have been able to point out 
the parametric relationship with the brane configurations of ABJM. We  
then compactified our theory on $CP^3$ and tried to find a relationship
 with the 
corresponding $CP^3$ compactification of the Galilean solution of the 
massive type IIA string theory \cite{hs}.   
We would like to conclude that the two type IIA theories compactified 
over $CP^3$,  the massive one  and 
the massless one with 2-form flux over $CP^3$, appear to be  the same  
in the non-relativistic case.  In the relativistic ABJM scenario, similar 
comparisons 
were studied in \cite{gaitt,gaitt2} involving  $CP^3$ compactifications. 

Our map seems to be true for the Galilean backgrounds with dynamical 
exponent  being 3 
and obviously without supersymmetry. Although the respective 10-dimensional 
solutions make distinct backgrounds but those appear to be related via 
our map in four dimensions. The global symmetries shared by these solutions 
are also found to be the same. 
We do not know what would be the 
situation with other such
non-relativistic solutions in the two theories. 
Particularly, the case of Galilean 
solutions with dynamical exponent 
2 would be interesting as those correspond  to  conformal Galilean 
CFTs.

\noindent{\it Acknowledgements:}\\
I am thankful to A. Dhar, S. 
Govindarajan and G. Mandal  for  discussions.



\end{document}